\documentclass{article}

\usepackage{PRIMEarxiv}

\usepackage[utf8]{inputenc} 
\usepackage[T1]{fontenc}    
\usepackage{hyperref}       
\usepackage{url}            
\usepackage{booktabs}       
\usepackage{amsfonts}       
\usepackage{nicefrac}       
\usepackage{microtype}      
\usepackage{lipsum}
\usepackage{fancyhdr}       
\usepackage{graphicx}       
\graphicspath{{media/}}     

\usepackage{titlesec}

\titlespacing*{\section}
{0pt}        
{1.2ex}      
{0.8ex}      

\title{Sola-Visibility-ISPM: \\ Benchmarking Agentic AI for Identity Security Posture Management Visibility
}

\author{
  \textbf{Gal Engelberg},
  \textbf{Konstantin Koutsyi},
  \textbf{Leon Goldberg},
  \textbf{Reuven Elezra},
  \textbf{Idan Pinto}\\
  \textbf{Tal Moalem},
  \textbf{Shmuel Cohen},
  \textbf{Yoni Weintrob}\\
  Sola Security, Tel Aviv, Israel \\
  \texttt{Corresponding author: gal.e@sola.security}
}

\begin{document}
\maketitle

\begin{abstract}

Identity Security Posture Management (ISPM) is a core challenge for modern enterprises operating across cloud and SaaS environments. Answering basic ISPM visibility questions, such as understanding identity inventory and configuration hygiene, requires interpreting complex identity data, motivating growing interest in agentic AI systems. Despite this interest, there is currently no standardized way to evaluate how well such systems perform ISPM visibility tasks on real enterprise data. We introduce the \textit{Sola Visibility ISPM Benchmark}, the first benchmark designed to evaluate agentic AI systems on foundational ISPM visibility tasks using a live, production-grade identity environment spanning AWS, Okta, and Google Workspace. The benchmark focuses on identity inventory and hygiene questions and is accompanied by the \textit{Sola AI Agent}, a tool-using agent that translates natural-language queries into executable data exploration steps and produces verifiable, evidence-backed answers. Across 77 benchmark questions, the agent achieves strong overall performance, with an expert accuracy of 0.84 and a strict success rate of 0.77. Performance is highest on AWS hygiene tasks, where expert accuracy reaches 0.94, while results on Google Workspace and Okta hygiene tasks are more moderate, yet competitive. Overall, this work provides a practical and reproducible benchmark for evaluating agentic AI systems in identity security and establishes a foundation for future ISPM benchmarks covering more advanced identity analysis and governance tasks.

\end{abstract}


\section{Introduction}\label{sec:intro}
Identity now defines the practical security boundary of the modern enterprise. As organizations operate across multi-cloud, SaaS, and hybrid environments, the integrity of identities, entitlements, authentication methods, and access paths determines the real perimeter of control. Industry guidance, such as RSA’s ISPM Framework \cite{RSA2023_ISPM} - highlights that identity misconfigurations, excessive privileges, unmanaged lifecycle states, and weak authentication posture are at the core of many security failures. ISPM has therefore become a foundational discipline for modern cybersecurity programs.

At the same time, agentic AI is reshaping how security operations are performed. Google Cloud’s 2025 Return on AI in Security report \cite{GoogleCloud2025_ROI_AI_Security} shows that organizations expect their largest operational gains from AI systems capable of autonomous reasoning, cross-system understanding, and end-to-end workflow execution. These systems can interpret complex telemetry, propose decisions, and accelerate response, capabilities that align directly with the data-intensive, multi-step nature of ISPM. As a result, agentic AI is increasingly viewed as a powerful accelerator for identity-centric security operations.

Yet despite this convergence, a significant gap remains: no standardized benchmark exists to evaluate how well agentic AI systems perform ISPM tasks, particularly in the context of question-answering agents operating over enterprise-grade identity and access data sources, a setup analogous to the data-grounded evaluation paradigm introduced in Spider-2.0 for SQL reasoning~\cite{Yao2023_Spider2_Text_to_SQL}. Existing cybersecurity benchmarks have advanced AI evaluation across several domains: interactive SOC investigation workflows \cite{ExCyTInBench_SecRL}, applied SOC reasoning over incident reports \cite{CyberSOCEval_CrowdStrike}, cyber threat intelligence reasoning \cite{CTIBench,SEvenLLM_Bench}, adversarial validation and robustness evaluation \cite{XBOW_Validation_Bench}, foundational NLP tasks in cybersecurity \cite{CyberBench_JPMC,CyberMetric}, ICS-focused knowledge evaluation \cite{SECURE_ICS_Knowledge_Eval}, vulnerability detection in code \cite{SecLLMHolmes_AI4CloudOps}, secured code generation \cite{Li2025GenSIaC}, and RBAC rule-following in synthetic access-control hierarchies \cite{OrgAccess_Benchmark_RBAC_Reasoning}.

However, none of these benchmarks evaluate the core tasks required for identity security posture management, such as interpreting identity inventories, parsing entitlements and privileges, validating lifecycle states, assessing configuration hygiene, performing cross-platform identity correlation, or generating posture-aware answers grounded in real organizational datasets. In other words, while the broader cybersecurity community has made progress benchmarking AI for investigations, CTI reasoning, and offensive capabilities, identity security, the control plane responsible for many modern breaches, remains unmeasured in the context of agentic AI systems reasoning over real identity data.

We outline nine operational ISPM dimensions that together represent the full lifecycle of identity security posture management, and describe the capabilities an agentic AI system must demonstrate within each:

\begin{itemize}
    \item \textbf{Visibility \& Hygiene} – Maintain complete, accurate identity inventories; detect misconfigurations and hygiene drift; evaluate MFA posture; and surface risky or stale identities.

    \item \textbf{Cross-System Correlation} – Connect identity signals across IdPs, cloud IAM, productivity suites, directories, and application layers; interpret federated trust; and reason end-to-end about identity exposure.

    \item \textbf{Behavioral Analytics} – Interpret authentication logs, privilege-use patterns, and other audit signals to detect suspicious or anomalous behavior.

    \item \textbf{Risk Assessment \& Scoring} – Rank identity risks based on posture, privilege, behavior, and potential business impact; identify identities that disproportionately increase organizational exposure.

    \item \textbf{Mitigation \& Recommendations} – Produce actionable, least-privilege-aligned remediation steps that consider operational constraints and security governance.

    \item \textbf{Framework Alignment \& Governance} – Map findings to standards such as NIST \cite{nist_sp_800_53r5}, CIS \cite{cis_controls_v8}, and ISO \cite{iso_iec_27001_2022}, and answer framework-level questions directly (e.g., ``Am I compliant with the CIS AWS Benchmark?''). Support audit, control validation, and policy-aligned reporting.

    \item \textbf{Contextual Threat Awareness} – Interpret identity exposures using known adversarial techniques, threat trends, and attack paths, and enrich reasoning with signals from CTI feeds to prioritize issues based on active or relevant threats.

    \item \textbf{Organizational Context Awareness} – Incorporate regulatory requirements, business-critical roles, sensitivity of assets, and organizational structures into reasoning.

    \item \textbf{Advanced Analytics} – Apply graph analysis, anomaly detection, multi-signal correlation, and other analytical techniques to detect emergent or latent identity risks beyond rule-based logic.
\end{itemize}

Across these dimensions, an agentic AI system must not only answer identity-security questions accurately, it must demonstrate correct data usage, multi-step reasoning, prioritization, and governance alignment consistent with real-world ISPM workflows.

In this paper, we introduce the Sola Visibility ISPM Benchmark, a focused benchmark covering the foundational ISPM tasks of \textbf{identity inventory, hygiene and misconfiguration detection}. It provides a curated set of security questions, and reproducible evaluation metrics for assessing agentic ISPM performance across environments built on AWS~\footnote{\url{https://aws.amazon.com/what-is-aws/}}, Okta~\footnote{\url{https://www.okta.com/products/workforce-identity/}}, and Google Workspace~\footnote{\url{https://workspace.google.com/}}. This benchmark establishes the core evaluation layer and lays the foundation for a broader, multi-dimensional Agentic ISPM Benchmark suite.

\section{Related Work}

AI evaluation in security and data systems has expanded significantly, driven by the need to measure models' reasoning ability, operational reliability, and alignment with real-world investigative or analytical workflows. Two major lines of work dominate this landscape: (1) text-to-SQL benchmarks that assess data-centric reasoning, and (2) cybersecurity-oriented benchmarks that evaluate threat analysis, SOC investigations, CTI reasoning, and vulnerability detection.

In the data systems domain, the Spider benchmarks provide the canonical foundation for evaluating natural-language interfaces to structured databases. Spider~1.0 \cite{yu2018spider} introduced cross-domain text-to-SQL evaluation through a static schema–query paradigm, pairing natural-language questions with human-authored SQL across 200 heterogeneous databases. Its execution-focused metrics emphasize syntactic correctness and semantic fidelity of generated SQL. Spider~2.0 \cite{Yao2023_Spider2_Text_to_SQL} advances this design by embedding models within realistic enterprise data environments such as BigQuery \cite{GoogleBigQuery} and Snowflake \cite{Snowflake}, requiring them to navigate multi-step workflows, revise errors, and reason over dependencies between code, schemas, and project context. This shift marks a broader methodological transition: from evaluating the correctness of a single output to evaluating the robustness and repairability of agentic workflows in complex data ecosystems.

Cybersecurity-focused benchmarks, in contrast, prioritize investigative reasoning, threat understanding, and vulnerability assessment. ExCyTIn-Bench \cite{ExCyTInBench_SecRL} places models within a simulated Azure SOC populated with live-style Sentinel telemetry \cite{AzureSentinel}, rewarding step-by-step investigative actions aligned with a human-designed threat investigation graph. This process-oriented feedback reflects the iterative nature of real SOC analysis, where intermediate hypotheses and investigative pivots matter as much as final conclusions. CyberSOCEval \cite{CyberSOCEval_CrowdStrike} complements this by grounding evaluation in authentic industry artifacts, CrowdStrike \cite{CrowdStrike} threat reports and sandbox analysis outputs, testing a model’s ability to extract, synthesize, and interpret operational threat intelligence directly from noisy real-world sources.

Cyber threat intelligence (CTI) reasoning benchmarks further probe a model’s ability to integrate structured vulnerability data with narrative threat reporting. CTIBench \cite{CTIBench} mixes expert-validated answers with CVE/CWE mappings and RAG-generated questions, enabling evaluation across both foundational knowledge and applied CTI reasoning. SEvenLLM-Bench \cite{SEvenLLM_Bench} extends CTI evaluation into a bilingual setting, introducing English–Chinese parallel corpora and hybrid metrics, including LLM-assisted semantic scoring, that more faithfully capture nuance in multi-language threat intelligence workflows.

Vulnerability detection is addressed by SecLLMHolmes \cite{SecLLMHolmes_AI4CloudOps}, which benchmarks static code analysis across C/C++ and Python codebases labeled with ground-truth vulnerabilities. Its multi-dimensional design tests whether models can robustly identify subtle security flaws across diverse code structures, bridging AI code understanding with practical vulnerability analysis.

General-purpose cybersecurity reasoning frameworks also contribute to this space. CyberBench \cite{CyberBench_JPMC} provides ten datasets spanning classification, summarization, name entity recognition (NER), and question answering (Q\&A), offering broad coverage of NLP skills commonly required in security workflows. CyberMetric \cite{CyberMetric} introduces a large-scale, expert-validated corpus designed to measure factual grounding and conceptual reasoning at certification-level depth. These benchmarks offer a complementary perspective to more interactive SOC or CTI settings, focusing on the foundational language and knowledge capabilities that support higher-level security tasks.

Identity and access management (IAM) reasoning, which plays a central role in modern security posture, has recently gained attention through OrgAccess \cite{OrgAccess_Benchmark_RBAC_Reasoning}. By constructing synthetic organizational hierarchies, permission sets, and RBAC rules, the benchmark evaluates whether models can interpret and apply structured access-governance policies. This line of work is increasingly relevant as enterprises adopt AI assistants to support identity governance, access reviews, and privilege analysis.
Unlike OrgAccess, which evaluates a model’s ability to interpret and apply predefined RBAC rules within synthetic organizational structures, our benchmark focuses on operational visibility into identity inventories and configuration hygiene in real-world enterprise environments. It assesses whether models can accurately enumerate identities, identify misconfigurations, and evaluate authentication posture grounded in production IAM, IdP, and SaaS data.

Overall, these developments highlight a clear shift toward multidimensional, context-aware evaluation: benchmarks are
moving beyond static accuracy to assess how reliably and realistically AI systems reason within security-critical and
data-intensive environments. This trend underscores the need for frameworks that capture not just whether a model is
correct, but whether it behaves robustly, and usefully in real operational settings.

\section{Sola AI Agent}
\label{sec:agent}

\begin{figure*}[ht]
    \centering
    \includegraphics[width=0.8\textwidth]{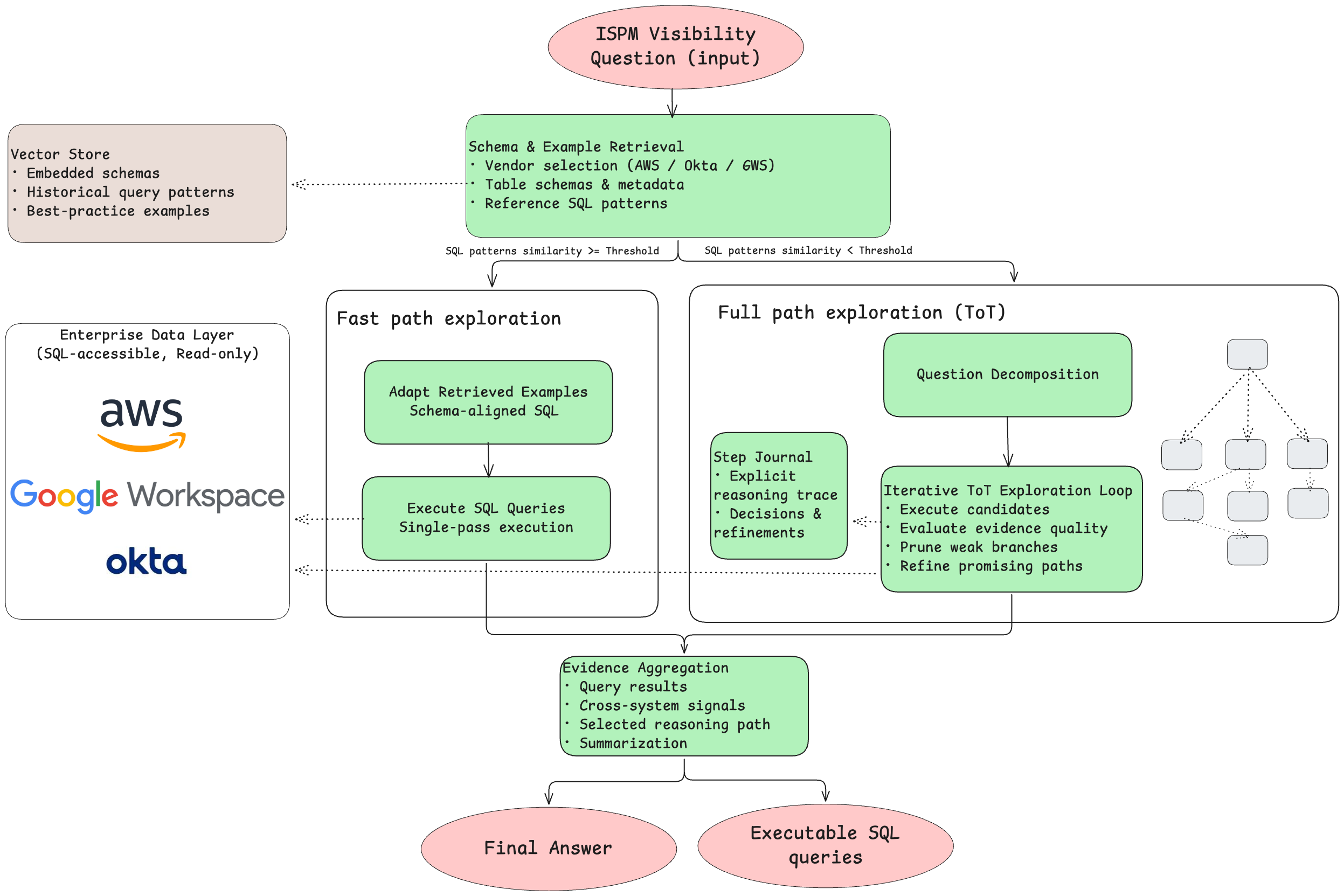}
    \caption{SOLA AI Agent Conceptual Architecture}
    \label{fig:sola-ispm-agent}
\end{figure*}

The Sola AI Agent is a tool-using, data-grounded security AI agent designed to answer complex security questions over enterprise data. While the agent itself is security-domain–generic, this work focuses on its application to Identity Security Posture Management (ISPM), where it is instantiated to address identity inventory and posture visibility questions across enterprise identity and access management environments.
Given a natural-language ISPM query, the agent translates the request into executable data exploration steps over production IAM, IdP, and SaaS tables, and returns a verifiable, evidence-backed answer. The agent follows a schema-grounded execution model (Figure~\ref{fig:sola-ispm-agent}). For each request, it first identifies the relevant identity platforms and retrieves the corresponding data schemas and reference query patterns. This grounding step constrains subsequent actions to valid tables, fields, and relationships, ensuring that all generated queries are executable and data-aligned. To balance efficiency and robustness, the agent supports two complementary execution modes: fast-path exploration and full-path exploration.

In fast-path exploration, the agent directly adapts retrieved example queries to the target schema and executes them in a single pass. This mode is used when example similarity and schema confidence are high, allowing low-latency enumeration of identity inventories and hygiene conditions without maintaining an explicit reasoning trace. Fast-path execution prioritizes efficiency while remaining fully grounded in underlying data.


In full-path exploration, the agent performs an explicit, iterative reasoning process. 
The question is decomposed into intermediate exploration steps, each associated with a concrete success criterion. 
The agent executes queries incrementally, validates intermediate results, and records its actions in a structured step journal.
This execution pattern is inspired by Tree-of-Thought–style reasoning \cite{yao2023tree}, in which intermediate states are explicitly evaluated and refined, but is grounded in executable data exploration steps rather than abstract reasoning alone.

Both execution modes converge on a shared evidence aggregation phase. Query results from all relevant platforms are consolidated and summarized into a final response that includes the natural-language answer, the executed queries, and the supporting evidence. When full-path execution is used, the complete step journal is retained, enabling inspection of the agent’s reasoning and decision-making process.

\section{Sola ISPM Visibility Benchmark}

\subsection{Data Source Integration}\label{subsection:datasource}

The Sola Visibility ISPM Benchmark is grounded on a \textit{live, production-grade enterprise environment} rather than synthetic or laboratory-generated setup. It includes operational identity and configuration data sources from three complementary platforms that collectively represent the technological foundation of cloud-centric organizations: an external identity provider (IdP), a cloud service provider (CSP), and a collaborative productivity platform. This tri-layer architecture reflects widely adopted enterprise identity ecosystems and enables realistic evaluation of identity-security reasoning.

First, an enterprise identity provider, implemented using Okta, supplies the authoritative workforce identity layer, including authentication flows, multi-factor authentication (MFA) policies, lifecycle management, and role-based access to organizational applications. Second, Amazon Web Services (AWS) serves as the cloud execution environment that hosts the organization’s operational workloads. The integrated dataset captures cloud identities, access keys, IAM roles, privilege assignments, and resource configurations. Third, Google Workspace (GWS) represents the organization’s collaborative and resource-governance layer. This includes workforce accounts, groups, Drive file permissions, OAuth application configurations, and administrative security controls.

Integrating these three operational platforms yields a unified, realistic identity environment in which all benchmark questions must be answered. Each question is constrained to be \emph{data-grounded}, meaning that responses must be derivable strictly from the configurations, relationships, and policies present in the environment.

\subsection{ISPM Questions Elicitation}

To construct a comprehensive and operationally grounded set of Identity Security Posture Management (ISPM) questions, we employed a structured elicitation methodology that systematically transforms authoritative security best practices into data-answerable natural language queries. This approach ensures that every benchmark question corresponds to a real configuration, policy, or identity relationship observable within the integrated enterprise environment.

The elicitation process draws from three families of best-practice sources, each corresponding directly to one of the platforms in the unified identity environment. From AWS, we used identity- and configuration-focused rules provided by the Scout Suite auditing framework \cite{scoutsuite}. From Google Workspace, we incorporated findings from publicly available security analyses, including the ScubaGoggles (\textit{Scube}) project \cite{scubagoggles}, which enumerates concrete Workspace misconfigurations and OAuth governance risks. From Okta, we relied on Okta’s published security best practices and identity-governance guidelines \cite{okta_security_best_practices}, which outline recommendations for MFA enforcement, session management, and privileged-access controls. These sources collectively provide the widely adopted practitioner baselines that informed the formulation of the benchmark’s ISPM questions.

Each best-practice rule was reformulated into a \emph{data-bounded} ISPM question, that is, a question whose correctness must be determined strictly from the data available within the integrated environment. Reformulation requires translating an abstract rule into a concrete, evaluable condition. For example, the Scout Suite rule \texttt{iam-managed-policy-allows-full-privileges}\cite{scoutsuite_allows_full_privileges}, which flags customer-managed IAM policies granting unrestricted privileges, was translated into the natural-language question: \emph{``Are there any managed policies that allow \texttt{Action=*} and \texttt{Resource=*}?''} Similarly, the rule \texttt{iam-managed-policy-no-attachments}\cite{scoutsuite_no_attachments}, identifying unused customer-managed IAM policies, was reformulated as: \emph{``Are there customer-managed policies with no attachments?''} These transformations ensure direct alignment between best-practice logic and the evidence available in the integrated AWS, Google Workspace, and Okta datasets.

Initial question candidates were generated using a large language model (GPT-5\cite{openai_gpt5_2025}), which was prompted to convert rule definitions into precise, operationally meaningful questions. All automatically generated questions underwent a multi-stage expert validation process. A panel of cybersecurity practitioners evaluated each question for semantic correctness and clarity. Questions whose semantics could not be fully validated through the available data sources, or whose answer could not be determined without external contextual knowledge, were refined or removed.

The resulting benchmark question-set\footnote{Full list available in Appendix \ref{app:sample-questions}} captures a representative identity-security visibility requirements across AWS, Google Workspace, and Okta. Because each question is directly derived from the real configurations and posture of these platforms, the benchmark provides meaningful coverage of identity-centric risks and enables rigorous evaluation of an agent’s data grounding, SQL reasoning, and cross-platform visibility.

\subsection{Evaluation Framework}

\begin{figure}[ht]
    \centering
    \includegraphics[width=0.8\textwidth]{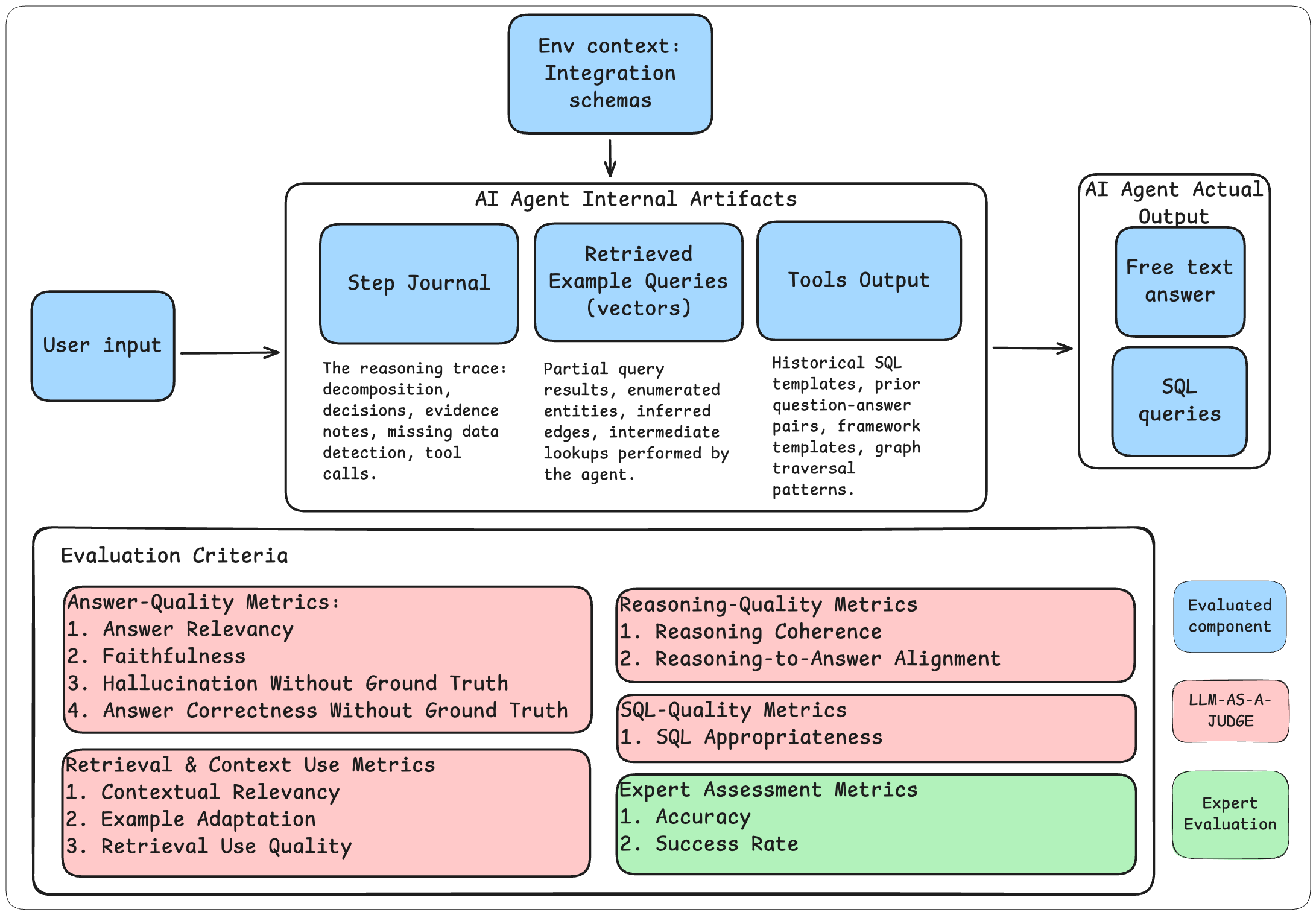}
    \caption{Sola ISPM Visibility Evaluation Framework}
    \label{fig:framework}
\end{figure}

Modern agentic AI systems for enterprise data question answering share a common architectural pattern composed of several interconnected components: a natural-language interpretation layer that parses the user query, a retrieval layer that locates relevant schema elements, examples, or historical traces, a reasoning engine that constructs a multi-step plan, a tool-execution layer responsible for generating and running SQL or API calls, and a synthesis layer that integrates intermediate results into a final natural-language answer. Although implementations differ, these components form the backbone of common agentic systems, and each introduces opportunities for both correct and faulty behavior. A comprehensive evaluation must therefore examine not only the final answer, but also the internal reasoning and the evidence used at each stage of the computation. To evaluate these components in a controlled and reproducible manner, We introduce a four-stage process.

\paragraph{Step 1: Executing All Questions Through the Sola Agent.}
After curating the benchmark question set, every question is executed end-to-end through the Sola agent. This execution yields a complete agentic reasoning trace, including the natural-language interpretation of the query, the reasoning steps taken, the generated SQL queries, and the results obtained from AWS, Google Workspace, and Okta. This approach mirrors the data-grounded evaluation paradigm introduced in data-aware text-to-SQL benchmarks such as Spider~2.0 \cite{Yao2023_Spider2_Text_to_SQL}, ensuring that both the reasoning path and the outputs are fully observable and assessable.

\paragraph{Step 2: Collection of the Evidence Bundle.}
For each question, we collect a structured set of evaluation artifacts that represent the behavior of different components of the agentic pipeline: \textit{Input}: the original ISPM question posed to the agent. \textit{Output (Natural Language)}: the final answer produced by the synthesis layer. \textit{Generated SQL}: the queries constructed by the tool-execution layer to retrieve the final answer. \textit{Retrieved Example Queries}: embeddings and examples surfaced by the retrieval layer, representing schema hints, patterns, and prior examples available to the agent. \textit{Step Journal and Tools Output}: a detailed, step-by-step record of the agent’s internal reasoning process, reflecting plan construction, schema interpretation, query formulation, synthesis logic, and collected evidence from executing SQL queries against AWS, GWS, and Okta. Each artifact provides insight into a different subsystem of the agent. Together, they form the evidence bundle used for expert evaluation and LLM-based judgment.

\paragraph{Step 3: Expert Evaluation.}
A panel of five cybersecurity experts then evaluates the outputs associated with each question. Experts assess the natural-language answers for semantic correctness and accurate interpretation of identity posture. Each answer and SQL artifact is assigned a correctness label of 0 (“Does Not Meet Criterion”), 0.5 (“Partially Meets Criterion”), or 1 (“Fully Meets Criterion”). Using these labels, the benchmark computes two deterministic metrics: \textit{Accuracy} and \textit{Success Rate}. Accuracy reflects the degree of correctness. It is the average of the expert-assigned scores and captures partial correctness where appropriate - for example, SQL that retrieves some but not all expected identities, or an answer that identifies the right misconfiguration but misses contextual details. Success Rate reflects strict correctness. It measures the percentage of examples where the model achieved a score of 1. Partial correctness (0.5) is not counted as success. This metric aligns with execution-style evaluation in modern text-to-SQL benchmarks\cite{Yao2023_Spider2_Text_to_SQL} and highlights the system’s ability to produce fully correct ISPM reasoning and SQL.

\paragraph{Step 4: LLM-as-Judge Evaluation.}
In parallel, we conduct LLM-as-Judge evaluation pipeline, applying structured chain-of-thought (CoT)\cite{Wei2022_Chain_of_Thought} scoring across a three-level rubric: 0 (“Does Not Meet Criterion”), 0.5 (“Partially Meets Criterion”), and 1 (“Fully Meets Criterion”). This approach follows the rubric-guided evaluation paradigm introduced by G-Eval\cite{Liu2023_GEVAL}, where LLM judges reason explicitly before issuing a score. Each benchmark instance is evaluated by a panel of two judges - Anthropic Claude Sonnet 4.5\cite{anthropic_claude_sonnet_4_5} and OpenAI GPT-4.1\cite{openai_gpt_4_1}, selected for complementary strengths. Sonnet 4.5 provides strong long-context and schema-aware reasoning, making it effective for evaluating step journals and multi-hop logic. GPT-4.1 provides semantically precise judgments and stable Q\&A performance under moderate token constraints. Combining these two evaluators yields balanced judgments: Sonnet excels at reasoning fidelity, while GPT-4.1 excels at semantic and factual precision. To maintain a conservative approach, the evaluated rank for each instance is determined by the minimum rank assigned by the two judges.
Evaluation is conducted across four components: (1) the user’s question, (2) the agent’s internal artifacts (step journal, retrieved examples, tool outputs), (3) the agent’s actual outputs (free-text and SQL), and (4) the expert-validated ground truth. Metrics are organized into evaluation criteria visible in Figure \ref{fig:framework}. A complete description of all evaluation metrics, including the G-Eval prompt formulations and rubric specifications, is available in Appendix~\ref{appendix:evaluation}.

\begin{itemize}
    \item \textbf{Answer-Quality:} Evaluates the final natural-language response. The \textit{AnswerRelevancy} judges whether the answer directly addresses the question’s intent. The \textit{Faithfulness} assesses whether the answer is grounded in schemas, step-journal evidence, and retrieved examples. The \textit{HallucinationNoGT} identifies invented tenant-specific claims or schema-invalid assertions. The \textit{AnswerCorrectnessNoGT} evaluates the likelihood that an answer is correct using SQL, reasoning steps, and schema structure, even in the absence of explicit ground truth. These metrics align with evidence-grounded Q\&A evaluation approaches emerging in retrieval-augmented LLMs and with recent work on grounding and answer faithfulness \cite{Sharma2023_RAGAS}.
    
    \item \textbf{Reasoning-Quality:} Assesses the step journal - the chain of reasoning that produced the answer. The \textit{ReasoningCoherence} measures whether reasoning is logically ordered, structurally sound, and consistent with the underlying schema. The \textit{ReasoningAnswerAlignment} evaluates whether the final answer logically follows from the reasoning trace. This approach is supported by recent work demonstrating that directly evaluating chain-of-thought improves reliability and interpretability \cite{Xu2025_DRO_Reasoning}.
    
    \item \textbf{Retrieval and Context-Use:} Evaluates how effectively the agent incorporates retrieved SQL templates, patterns, and factual evidence from tool outputs. The \textit{ContextualRelevancy} checks whether retrieved examples and evidence are relevant to the question. The \textit{RetrievalUseQuality} evaluates whether schemas and examples are actually used to construct better SQL and reasoning. The \textit{ExampleAdaptation} measures whether example SQL templates are adapted - not copied - to match the current schema and question. These metrics are informed by retrieval-augmented evaluation frameworks such as RAGAS \cite{Sharma2023_RAGAS} and by recent studies on retrieval precision and RAG robustness \cite{yu2024evaluation}.

    \item \textbf{SQL-Quality:} Evaluates whether the generated SQL is structurally appropriate, schema-aligned, and logically capable of solving the question. \textit{SQLSemanticAppropriateness} checks whether the SQL aligns with the schema and logically answers the question. This structural SQL evaluation follow the methodology established in Spider 2.0 and state of the art text-to-SQL analyses \cite{Yao2023_Spider2_Text_to_SQL,Min2023_Benchmarking_Text_to_SQL_LLMs}, which emphasize schema alignment, structural validity, and operator-level correctness independent of ground-truth SQL references.
\end{itemize}

\section{Experiment Results}
\label{sec:results}

This section presents the empirical evaluation results of the Sola agent using the Sola Visibility ISPM Benchmark questions over the production grade environment descibed in subsection \ref{subsection:datasource}. We first report overall correctness across ISPM domains, combining LLM-as-Judge signals with expert-based assessment. We then analyze performance by reasoning strategy, distinguishing between full-path and fast-path reasoning. Together, these results provide a structured view of agentic AI performance for ISPM visibility tasks and highlight systematic strengths and limitations across domains and reasoning modes.

It is important to note that all evaluation measures in this benchmark, including expert-based accuracy and LLM-as-Judge metrics, are computed over an ordinal scoring scale \{0, 0.5, 1\}, corresponding to incorrect, partially correct, and fully correct outcomes, respectively. Consequently, reported metric values should not be interpreted as binary or threshold-based scores. For example, a correctness or accuracy score of 0.6 indicates that, on average, system outputs fall between partially and fully correct, rather than reflecting a low rate of correct responses. This ordinal formulation is intentionally designed to capture graded performance in ISPM visibility tasks, where answers may be substantively correct while missing secondary contextual details.

\paragraph{Overall Correctness Across ISPM Domains.}


Table~\ref{tab:overall-correctness} summarizes overall correctness across ISPM domains using three complementary indicators: AnswerCorrectnessNoGT, an LLM-as-judge metric; expert-based Accuracy, which captures graded correctness and allows partial credit; and expert-based Success Rate, which measures strict correctness and requires fully correct answers. Overall, the benchmark comprises 77 questions and demonstrates strong performance across all metrics, with an AnswerCorrectnessNoGT score of 0.82, expert accuracy of 0.84, and a success rate of 0.77.

At the domain level, Sola achieves its highest correctness on AWS Hygiene questions, with an expert accuracy of 0.95 and a success rate of 0.90, indicating that the vast majority of responses are fully correct. Inventory questions also show solid performance, with an expert accuracy of 0.75 and a success rate of 0.64, reflecting a combination of fully correct and partially correct answers. Similarly, Google Workspace Hygiene exhibits strong correctness, with an expert accuracy of 0.75 and a success rate of 0.71. In contrast, Okta Hygiene shows more moderate performance: while expert accuracy reaches 0.65, the success rate drops to 0.50, suggesting that partially correct answers are more common than fully correct ones in this domain.

\begin{table}[ht]
\centering
\small
\begin{tabular}{lccccc}
\hline
\textbf{Measure / ISPM Domain} & \textbf{AWS} & \textbf{GWS} & \textbf{Inventory} & \textbf{Okta} & \textbf{Total} \\
\hline
\# Questions & 39 & 14 & 14 & 10  & 77 \\
LLM-as-Judge (AnswerCorrectnessNoGT) & 0.92 & 0.54 & 0.84 & 0.83  & 0.82 \\
Expert Accuracy & 0.95 & 0.75 & 0.75 & 0.65 & 0.84 \\
Expert Success Rate & 0.90 & 0.71 & 0.64 & 0.50 & 0.77 \\
\hline
\end{tabular}
\caption{Overall correctness by ISPM domain.}
\label{tab:overall-correctness}
\end{table}

\paragraph{Performance by Full-Path Reasoning.}

Full-path reasoning (Table~\ref{tab:full-path}) demonstrates strong overall performance, achieving an AnswerCorrectnessNoGT score of 0.87, an expert accuracy of 0.81, and an expert success rate of 0.75 across 40 questions. Correctness is highest for AWS Hygiene and Okta Hygiene, where both automated and expert-based metrics indicate consistently strong performance. In contrast, Google Workspace Hygiene exhibits substantially lower correctness, with an AnswerCorrectnessNoGT of 0.53 and correspondingly lower expert accuracy and success rate (0.65 and 0.6 respectively).

Beyond correctness, full-path execution shows consistently high answer and reasoning quality across domains. Answer relevancy and faithfulness remain near-perfect overall, while reasoning coherence and reasoning–answer alignment are uniformly strong, with perfect alignment observed across all domains. These results indicate that when answers are produced, they are well-supported by coherent reasoning and internally consistent explanations.

Analysis of the execution components further suggests that correct answers are typically grounded in appropriate query construction and context usage. SQL semantic appropriateness and retrieval use quality remain high overall, particularly for AWS and Okta, indicating effective adaptation of retrieved schemas and examples.

A key factor underlying correctness differences across domains is the quality of example adaptation. We observe a moderate positive correlation between Example Adaptation and AnswerCorrectnessNoGT (Pearson correlation $\rho \approx 0.5$), suggesting that the agent’s ability to adapt retrieved SQL templates and patterns to the target schema is an important driver of correctness. This relationship persists even under full-path reasoning, indicating that high-quality example adaptation provides essential structural grounding that complements, rather than replaces, explicit multi-step reasoning.

\begin{table}[ht]
\centering
\begin{tabular}{lccccc}
\hline
\textbf{Measure / ISPM Domain} & \textbf{AWS} & \textbf{GWS} & \textbf{Inventory} & \textbf{Okta} & \textbf{Total} \\
\hline
\# Questions & 18 & 10 & 4 & 8 & 40 \\
\hline
Answer Relevancy & 1.00 & 1.00 & 0.90 & 1.00 & 0.99 \\
Faithfulness & 1.00 & 1.00 & 0.83 & 0.98 & 0.98 \\
AnswerCorrectnessNoGT & 1.00 & 0.53 & 0.88 & 0.98 & 0.87 \\
Contextual Relevancy & 0.94 & 0.64 & 0.70 & 0.80 & 0.81 \\
Reasoning Coherence & 0.97 & 0.70 & 0.78 & 1.00 & 0.89 \\
Reasoning--Answer Alignment & 1.00 & 1.00 & 1.00 & 1.00 & 1.00 \\
SQL Semantic Appropriateness & 0.97 & 0.62 & 0.83 & 0.88 & 0.85 \\
Retrieval Use Quality & 0.97 & 0.61 & 0.78 & 0.90 & 0.85 \\
HallucinationNoGT & 1.00 & 0.90 & 0.75 & 0.63 & 0.88 \\
Example Adaptation & 0.71 & 0.32 & 0.80 & 0.71 & 0.62 \\
\hline
Expert Accuracy & 0.94 & 0.65 & 0.75 & 0.75 & 0.81 \\
Expert Success Rate & 0.89 & 0.60 & 0.75 & 0.63 & 0.75 \\
\hline
\end{tabular}
\caption{Full-path reasoning results across all evaluation dimensions.}
\label{tab:full-path}
\end{table}

\paragraph{Performance by Fast-Path Reasoning.}



Fast-path reasoning (Table~\ref{tab:fast-path}) achieves strong overall performance, with an AnswerCorrectnessNoGT score of 0.78, an expert accuracy of 0.86, and an expert success rate of 0.78 across 37 questions. However, performance exhibits greater variance across ISPM domains compared to full-path execution. AWS Hygiene and Inventory remain robust under fast-path reasoning, with consistently high automated and expert-based correctness. In contrast, Okta Hygiene shows markedly lower correctness; however, this result is based on only two Okta samples routed to fast-path execution, limiting the statistical significance of this observation.

Because fast-path execution bypasses explicit step journals, metrics that depend on reasoning traces, such as faithfulness, reasoning coherence, and reasoning–answer alignment, are not applicable in this setting. Among the remaining evaluation signals, performance is primarily driven by contextual grounding and schema alignment. Contextual relevancy and retrieval use quality vary across domains, while SQL semantic appropriateness remains reasonable overall, indicating that answers are generally supported by structurally valid queries even in the absence of explicit reasoning traces.

Consistent with this behavior, we observe a strong positive correlation between example adaptation and answer correctness (Pearson correlation $\rho \approx 0.8$), suggesting that when fast-path reasoning is used, the ability to effectively adapt retrieved examples and SQL patterns plays a central role in determining correctness.

\begin{table}[ht]
\centering
\begin{tabular}{lccccc}
\hline
\textbf{Measure / ISPM Domain} & \textbf{AWS} & \textbf{GWS} & \textbf{Inventory} & \textbf{Okta} & \textbf{Total} \\
\hline
\# Questions & 21 & 4 & 10 & 2 & 37 \\
\hline
Answer Relevancy & 1.00 & 1.00 & 0.95 & 1.00 & 0.99 \\
AnswerCorrectnessNoGT & 0.84 & 0.58 & 0.83 & 0.25 & 0.78 \\
Contextual Relevancy & 0.80 & 0.43 & 0.73 & 0.50 & 0.73 \\
SQL Semantic Appropriateness & 0.72 & 0.70 & 0.85 & 0.90 & 0.76 \\
Retrieval Use Quality & 0.70 & 0.63 & 0.94 & 0.70 & 0.75 \\
HallucinationNoGT & 0.52 & 0.50 & 0.55 & 0.00 & 0.50 \\
Example Adaptation & 0.68 & 0.68 & 0.90 & 0.50 & 0.73 \\
\hline
Expert Accuracy & 0.95 & 1.00 & 0.75 & 0.25 & 0.86 \\
Expert Success Rate & 0.90 & 1.00 & 0.60 & 0.00 & 0.78 \\
\hline
\end{tabular}
\caption{Fast-path reasoning results. Reasoning-based metrics are not applicable due to the absence of step journals.}
\label{tab:fast-path}
\end{table}

\paragraph{Fast-Path vs. Full-Path Reasoning.}

Comparing fast-path and full-path execution reveals a trade-off between explicit reasoning and reliance on example grounding. Full-path reasoning achieves higher and more consistent correctness across domains, particularly for more complex hygiene settings such as Okta and Google Workspace, while fast-path reasoning performs well for AWS and Inventory but exhibits greater variability. Correctness under fast-path execution is more strongly coupled to example adaptation quality, as indicated by a higher correlation with AnswerCorrectnessNoGT ($\rho \approx 0.8$) than under full-path execution ($\rho \approx 0.5$), since by design fast-path relies more heavily on structurally well-aligned examples, whereas full-path reasoning can partially compensate for weaker example alignment through explicit multi-step reasoning.

\paragraph{Agreement Between Automated and Expert Evaluation.}

Table~\ref{tab:agreement} reports agreement between expert judgments and automated evaluation signals using \emph{mean absolute agreement (MAA)}, computed as the mean absolute difference between automated scores and expert assessments. Lower values indicate stronger agreement. Overall, automated evaluations exhibit reasonably close agreement with expert judgments, with a grand-total MAA of 0.22 for the aggregated LLM-as-Judge mean and 0.26 for the minimum aggregation.

\begin{table}[ht]
\centering
\small
\begin{tabular}{lcccc}
\hline
\textbf{ISPM Domain} &
\textbf{Sonnet 4.5} &
\textbf{GPT-4.1} &
\textbf{LLM-as-Judge (Mean)} &
\textbf{LLM-as-Judge (Min)} \\
\hline
AWS Hygiene        & 0.10 & 0.09 & 0.09 & 0.12 \\
GWS Hygiene        & 0.64 & 0.30 & 0.47 & 0.64 \\
Inventory          & 0.32 & 0.23 & 0.28 & 0.34 \\
Okta Hygiene       & 0.18 & 0.35 & 0.27 & 0.18 \\
\hline
\textbf{Grand Total} & \textbf{0.25} & \textbf{0.19} & \textbf{0.22} & \textbf{0.26} \\
\hline
\end{tabular}
\caption{Mean absolute agreement (MAA) between expert judgments and automated evaluation signals. Lower values indicate stronger agreement.}
\label{tab:agreement}
\end{table}

Agreement varies substantially across ISPM domains. AWS Hygiene shows the strongest agreement, with consistently low MAA values across all judges (0.09--0.12), reflecting near-alignment between automated and expert evaluations. Okta Hygiene also demonstrates relatively strong agreement overall, though with greater variability across individual judges. In contrast, Google Workspace Hygiene exhibits the weakest agreement, with substantially higher MAA values (up to 0.64), indicating larger and more frequent discrepancies between automated scoring and expert judgment. Inventory falls between these extremes, showing moderate agreement. Across domains, aggregated LLM-as-Judge scores exhibit agreement with expert assessments that is comparable to, but not uniformly better than, individual judges. In some domains, aggregation reduces variability across evaluators, while in others individual judges achieve lower mean absolute agreement.

\section{Conclusion}

This paper introduced the \textbf{Sola Visibility ISPM Benchmark}, the first benchmark specifically designed to evaluate agentic AI systems on foundational Identity Security Posture Management (ISPM) visibility tasks. By grounding evaluation in a live, production-grade identity environment spanning AWS, Okta, and Google Workspace, the benchmark captures the operational complexity of real enterprise identity systems and evaluates agentic behavior under realistic data, schema, and execution constraints.

Using a comprehensive evaluation framework that integrates expert assessment with structured LLM-as-Judge metrics, we demonstrate that agentic AI systems achieve \textbf{strong graded correctness} on ISPM visibility tasks, with particularly robust performance in AWS-centric environments. Because all evaluation metrics operate on an ordinal scale, reported accuracy values reflect degrees of partial to full correctness rather than binary success or failure. This design surfaces an important practical insight: even as task complexity increases, the agent frequently produces outputs that are structurally valid, evidence-backed, and operationally meaningful, despite occasionally incomplete coverage of secondary details.

Beyond final-answer correctness, our results provide visibility into how agentic workflows succeed or degrade. Answer relevancy remains consistently high across all domains, indicating that performance gaps rarely arise from misinterpretation of task intent. Instead, variations in performance are driven primarily by the quality of contextual grounding, schema-aligned SQL construction, and example adaptation. Across both full-path and fast-path execution modes, effective adaptation of retrieved query patterns emerges as a central driver of correctness. Notably, this effect persists even when explicit multi-step reasoning traces are available, suggesting that robust structural grounding and pattern adaptation often play a more cardinal role than the presence of detailed reasoning chains alone.

Taken together, these findings emphasize that reliable agentic performance in identity-security visibility tasks depends on the system’s ability to execute, adapt, and ground multi-step workflows over complex enterprise schemas and data, rather than on isolated reasoning depth or single-shot accuracy. The Sola Visibility ISPM Benchmark therefore establishes a rigorous and reproducible baseline for evaluating agentic AI systems in real-world identity-security environments. Future work will extend this benchmark into a broader ISPM evaluation suite, incorporating higher-order capabilities such as cross-system correlation, behavioral analytics, identity risk scoring, mitigation planning, governance alignment, and advanced graph-based identity reasoning.

\bibliographystyle{unsrt}  
\bibliography{references}

\appendix
\section{Benchmark Question Set}
\label{app:sample-questions}

This appendix lists the complete set of ISPM visibility questions used in the Sola Visibility ISPM Benchmark. The questions are organized by domain and platform and define the concrete, data-bounded tasks used to evaluate agentic AI systems for identity security posture visibility.

\paragraph{Inventory}

\begin{quote}
\vspace{-1em}
\textit{\begin{enumerate}
    \item List high-privilege human identities by type (AWS IAM, Okta).
    \item List high-privilege non-human identities by type (AWS execution roles, AWS cross-account roles).
    \item List secrets in AWS per resource and per owner type (root, IAM user, role).
    \item Which Google Workspace admin users are missing 2-Step Verification?
    \item List all Super Admins (for periodic review).
    \item Which IAM users have console passwords enabled but no MFA?
    \item Which active access keys are older than 90 days?
    \item Which IAM roles trust external accounts (not in AWS Organizations)?
    \item Which roles have inline policies attached?
    \item Is root MFA enforced?
    \item Which sign-on policies exceed a 120-minute session lifetime?
    \item Which users are active but have not signed in for over 90 days?
    \item Which applications’ default access rules do not require MFA?
    \item Which Okta connections lack signed and encrypted SAML assertions?
\end{enumerate}}
\end{quote}

\paragraph{AWS Hygiene}

\begin{quote}
\vspace{-1em}
\textit{\begin{enumerate}
    \item Do any IAM roles have trust policies that allow principals from other accounts without requiring MFA or an External ID?
    \item Do any IAM roles allow the sts:AssumeRole action without requiring MFA in their trust policies?
    \item Are there IAM roles whose trust policies allow any AWS principal (e.g., Principal set to “*”)?
    \item Are there EC2 service roles that are not attached to any running EC2 instances?
    \item Do any IAM groups have inline policies attached?
    \item Do any human IAM users have policies attached directly (inline or managed)?
    \item Do any inline policies in IAM identities use NotAction with an Allow effect?
    \item Do any user inline policies allow actions other than sts:AssumeRole?
    \item Do any IAM roles have inline policies that allow a specific service:action on all resources (*)?
    \item Do any policies deny user actions on group resources rather than on the users themselves?
    \item Do any customer-managed IAM policies use NotAction in Allow statements?
    \item Are there any managed policies that allow Action = "*" and Resource = "*"?
    \item Do any customer-managed policies grant permissions other than sts:AssumeRole?
    \item Do any customer-managed policies allow a specific service:action on all resources (*)?
    \item Are there customer-managed policies with no attachments?
    \item Is the AWSSupportAccess AWS managed policy attached to at least one principal (user, group, or role)?
    \item Do any IAM roles have inline policies attached?
    \item Is a hardware MFA device enabled on the root account?
    \item Is MFA enabled on the root account?
    \item Has the root account been used within the last 90 days?
    \item Does the root account have any active X.509 certificates?
    \item Does the root account have any active access keys?
    \item Do any service users (identified by membership in a service-user group) have console passwords enabled?
    \item Are there any users with passwords or access keys that have not been used in over 90 days and are still enabled?
    \item Do any active access keys have a creation date older than 90 days?
    \item Are there users who do not belong to the common IAM group?
    \item Are there users with access keys that were created during setup but never used?
    \item Do any users have more than one active access key?
    \item Do any IAM users have both a console password and one or more active API keys?
    \item Are there users with inline or managed policies attached directly?
    \item Do all password-enabled IAM users have MFA enabled?
    \item Does the account password policy enforce password expiration and set MaxPasswordAge $\leq$ 90 days?
    \item Is the IAM password policy configured to require at least 14 characters?
    \item Is password expiration enabled in the IAM password policy?
    \item Does the IAM password policy require at least one lowercase letter?
    \item Does the password policy require at least one numeric character?
    \item Does the password policy require at least one special character?
    \item Does the IAM password policy require at least one uppercase letter?
    \item Does the password policy prevent the reuse of previous passwords?
\end{enumerate}}
\end{quote}

\paragraph{Google Workspace Hygiene}

\begin{quote}
\vspace{-1em}
\textit{\begin{enumerate}
    \item Are there any users whose MFA method is configured as SMS or voice call?
    \item Is the new user 2-Step Verification enrollment period configured between 1 and 7 days for each organization unit?
    \item Are Google Workspace domains configured with Single Sign-On ensuring that post-SSO verification is enabled for the primary SSO profile?
    \item Is the minimum password length at least 12 characters?
    \item Is the minimum password length configured to 15 characters or more?
    \item Is the password policy configured to be enforced at the next sign-in for all organizational units?
    \item Is password reuse disabled for all organizational units?
    \item Are password expiration policies disabled for all organizational units?
    \item Do all privileged accounts (Super Admins and other high-privilege roles) authenticate with Google credentials and use phishing-resistant MFA?
    \item Is the number of Google Workspace Super Admin users between 2 and 8 inclusive?
    \item Are self-service account recovery options disabled for users assigned the Super Admin role?
    \item Are account recovery options disabled for all users who are not Super Admins?
    \item Are all users labeled as sensitive enrolled in Google’s Advanced Protection Program?
    \item Are all domains configured to restrict third-party applications and disallow less-secure apps?
\end{enumerate}}
\end{quote}

\paragraph{Okta Hygiene}

\begin{quote}
\vspace{-1em}
\textit{\begin{enumerate}
    \item Are there any sign-on policies where the maximum session lifetime (\texttt{max\_session\_lifetime\_minutes}) exceeds 120 minutes (2 hours)?
    \item Are MFA policies missing phishing-resistant authenticators?
    \item Are there any sign-on policies where \texttt{require\_mfa\_each\_signin} is FALSE or NULL?
    \item Which applications have \texttt{mfa\_required} set to FALSE or NULL?
    \item Which active users have not signed in for more than 90 days?
    \item Does the password policy enforce a minimum length of at least 12 characters and require uppercase letters, lowercase letters, numbers and symbols?
    \item Are any legacy authentication factors (SMS, voice, email, OTP) enabled?
    \item Which sign-on policies have a reauthentication interval greater than 480 minutes (8 hours) or have no reauthentication interval set?
    \item Are there any active users who are no longer members of any group assigned through an Application or IdP?
    \item Do any identity providers (IdPs) lack signed or encrypted SAML assertions?
\end{enumerate}}
\end{quote}




\section{Evaluation Criteria and Rubrics}
\label{appendix:evaluation}

This appendix provides the complete criteria prompts and rubrics used in the 
LLM-as-Judge evaluation of the SOLA Visibility ISPM Benchmark. For each metric, 
we include: (1) a short description of what is evaluated, (2) the exact criteria 
prompt provided to the judging model, and (3) the scoring rubric. All metrics 
use a three-level scale: 0 (Does Not Meet Criterion), 0.5 (Partially Meets 
Criterion), and 1 (Fully Meets Criterion).

\subsection{Answer Relevancy}

\paragraph{What is evaluated.}
This metric evaluates how directly and completely the agent's final answer addresses the
security question posed. In the context of Identity Security Posture Management (ISPM),
relevant answers must stay aligned with identity-centric risk, privilege posture, lifecycle
state, authentication strength, access paths, and cloud/SaaS configuration semantics.

\paragraph{Criteria (evaluation prompt).}
\begin{quote}\ttfamily\raggedright
You are a senior expert in the domain `Identity Security Posture Management (ISPM)`, 
specializing in Questions about identities (human and non-human), their privileges, 
access paths, authentication posture (MFA, phishing-resistant methods), lifecycle hygiene 
(onboarding/offboarding, dormant accounts), policy/configuration posture, and resulting 
identity-centric risk across cloud and SaaS environments.

You are evaluating how relevant a model's answer is to a security question.

Task:
- Evaluate how relevant the model's ANSWER is to the QUESTION in this domain.
- Consider whether the answer directly addresses the user's intent and stays on-topic.
\end{quote}

\paragraph{Rubric.}
\begin{quote}\ttfamily\raggedright
0.0 — Irrelevant:  
\quad does not address the question or is mostly off-topic.

0.5 — Partially relevant:  
\quad some on-topic content, but incomplete or mixed with irrelevant parts.

1.0 — Fully relevant:  
\quad directly addresses the main intent and stays on-topic.
\end{quote}

\subsection{Faithfulness}

\paragraph{What is evaluated.}
This metric assesses whether the agent's answer is fully grounded in the provided 
evidence: schemas, retrieved examples, and reasoning traces. In the ISPM domain, 
faithful answers must accurately align with identity structures, access relationships, 
configuration states, log-derived evidence, and schema-defined constraints.

\paragraph{Criteria (evaluation prompt).}
\begin{quote}\ttfamily\raggedright
You are a senior expert in the domain `Identity Security Posture Management (ISPM)`, 
specializing in Questions about identities (human and non-human), their privileges, 
access paths, authentication posture (MFA, phishing-resistant methods), lifecycle 
hygiene (onboarding/offboarding, dormant accounts), policy/configuration posture, 
and resulting identity-centric risk across cloud and SaaS environments.

You are evaluating whether an answer is faithful to the available evidence.

CONTEXT may include:
- SCHEMAS: database/log schemas that define valid structures.
- RETRIEVED\_EXAMPLES: historical questions/answers/SQL patterns from a helper 
  agent (PATTERNS, not tenant ground truth; may also appear inside STEP\_JOURNAL).
- STEP\_JOURNAL: reasoning and tool calls for this run, including factual evidence 
  (e.g., sample query results, discovered configs).

Use SCHEMAS as authoritative for structures. Use RETRIEVED\_EXAMPLES as 
method/pattern hints, not as strict ground truth. Judge whether the ANSWER's factual 
claims and structural references are supported or contradicted by SCHEMAS and any 
explicit evidence about the current environment.
\end{quote}

\paragraph{Rubric.}
\begin{quote}\ttfamily\raggedright
0.0 — Unfaithful:  
\quad major claims unsupported or contradicted by schema/evidence.

0.5 — Partially faithful:  
\quad some claims grounded, others speculative or weakly supported.

1.0 — Fully faithful:  
\quad important claims are supported and compatible with schema/evidence.
\end{quote}

\subsection{Hallucination (No Ground Truth)}

\paragraph{What is evaluated.}
This metric evaluates whether the agent introduces unsupported tenant-specific facts, 
schema-inconsistent entities, invented SQL structures, or fabricated evidentiary details.  
In Identity Security Posture Management (ISPM), hallucinations often manifest as invented 
IAM entities, nonexistent roles or permissions, false statements about authentication posture, 
or contradictions with audit/log evidence.

\paragraph{Criteria.}
\begin{quote}\ttfamily\raggedright
You are a senior expert in the domain `Identity Security Posture Management (ISPM)`, 
specializing in Questions about identities (human and non-human), their privileges, 
access paths, authentication posture (MFA, phishing-resistant methods), lifecycle hygiene 
(onboarding/offboarding, dormant accounts), policy/configuration posture, and resulting 
identity-centric risk across cloud and SaaS environments.

You are evaluating whether the agent's FINAL ANSWER contains hallucinations. IMPORTANT: 
You must evaluate hallucination across the entire pipeline (ANSWER + SQL + STEP\_JOURNAL), 
not only the answer text.

CONTEXT includes:
- SCHEMAS: authoritative tables, columns, and structural constraints.
- RETRIEVED\_EXAMPLES: historical patterns (NOT tenant ground truth).
- STEP\_JOURNAL: reasoning, tool calls, discovered evidence.
- MODEL\_SQL\_LIST: SQL generated by the agent.
- FINAL\_ANSWER: the answer being judged.

You must distinguish THREE types of hallucinations:

1. TENANT-SPECIFIC HALLUCINATION (fatal):
- Inventing tables/columns/fields not in the schema.
- Claiming tenant-specific facts NOT found in SQL/journal evidence.
- Referencing impossible configurations contradicted by schema.
These MUST be scored as 0.0.

2. PIPELINE HALLUCINATION (moderate):
- SQL uses non-existent tables/columns.
- STEP\_JOURNAL invents evidence or contradicts itself.
- FINAL\_ANSWER depends on hallucinated SQL or reasoning.
These should score 0.0 or 0.5 depending on severity.

3. DOMAIN SPECULATION (allowed but limited):
- Generic claims about security ('MFA reduces risk', 'privileged accounts are risky'),
  which do NOT pretend to be tenant-specific.
- These MUST NOT be treated as hallucinations unless incorrectly presented as tenant facts.
These may score 1.0 or 0.5 depending on alignment.

Evaluation logic:
- 1.0: No tenant-specific or pipeline hallucinations; domain-level statements are correct.
- 0.5: Minor speculative or generic claims OR mild misuse of examples, but no invented tenant facts.
- 0.0: Any tenant-specific hallucination OR SQL/journal-based hallucination contaminating the answer.
\end{quote}

\paragraph{Rubric.}
\begin{quote}\ttfamily\raggedright
0.0 — Hallucinated: invented tenant-specific facts OR invalid tables/columns OR SQL/reasoning 
hallucinations contaminating the answer.

0.5 — Some speculation or weakly supported statements, but no invented tenant facts; 
domain-level generalizations may be imperfect but not false.

1.0 — No hallucinations: answer and pipeline remain grounded in schema and evidence; 
any domain-level statements are correct and clearly not tenant-specific.
\end{quote}

\subsection{Answer Correctness vs Ground Truth}

\paragraph{What is evaluated.}
This metric determines whether the agent's final answer reaches the same security-relevant 
conclusions as the ground-truth answer, regardless of phrasing.  
In ISPM this includes privilege classification, lifecycle hygiene interpretation, exposure 
determination, authentication posture, and identity risk assessments.

\paragraph{Criteria.}
\begin{quote}\ttfamily\raggedright
You are a senior expert in the domain `Identity Security Posture Management (ISPM)`, 
specializing in Questions about identities (human and non-human), their privileges, 
access paths, authentication posture (MFA, phishing-resistant methods), lifecycle hygiene 
(onboarding/offboarding, dormant accounts), policy/configuration posture, and resulting 
identity-centric risk across cloud and SaaS environments.

You are comparing a model answer to a ground-truth answer.

ACTUAL\_OUTPUT is the model's answer. EXPECTED\_OUTPUT is the ground-truth answer.
Judge whether they express the same security-relevant conclusion (entities, classifications, 
posture, counts, etc.).
\end{quote}

\paragraph{Rubric.}
\begin{quote}\ttfamily\raggedright
0.0 — Incorrect: model answer reaches a different or wrong conclusion.

0.5 — Partially correct: some overlap with GT, but important differences or omissions.

1.0 — Correct: model answer is semantically equivalent to the ground truth.
\end{quote}

\subsection{Reasoning Coherence}

\paragraph{What is evaluated.}
This metric evaluates the internal logical consistency of the agent’s reasoning trace (`STEP\_JOURNAL`). A coherent ISPM reasoning trace should correctly reason about privileges, 
roles, access relationships, identity posture, configuration sources, and supporting evidence 
without contradictions or schema violations.

\paragraph{Criteria.}
\begin{quote}\ttfamily\raggedright
You are a senior expert in the domain `Identity Security Posture Management (ISPM)`, 
specializing in Questions about identities (human and non-human), their privileges, 
access paths, authentication posture (MFA, phishing-resistant methods), lifecycle hygiene 
(onboarding/offboarding, dormant accounts), policy/configuration posture, and resulting 
identity-centric risk across cloud and SaaS environments.

You are evaluating the coherence of a reasoning trace.

CONTEXT may include:
- SCHEMAS: for checking structural plausibility.
- RETRIEVED\_EXAMPLES: historical patterns that may also appear inside the journal.

The ACTUAL\_OUTPUT is the STEP\_JOURNAL (reasoning trace) for the given QUESTION.
Focus on logical order, internal consistency, and whether the reasoning steps make sense 
for the question and schema/domain. Penalize reasoning that relies on impossible tables/columns 
or that misuses examples in clearly inconsistent ways.
\end{quote}

\paragraph{Rubric.}
\begin{quote}\ttfamily\raggedright
0.0 — Incoherent: disorganized, contradictory, or clearly inconsistent with schema/domain.

0.5 — Partially coherent: some logical flow, but gaps, contradictions, or schema misuse.

1.0 — Coherent: clear, logically ordered, and consistent with schema/domain.
\end{quote}

\subsection{Reasoning–Answer Alignment}

\paragraph{What is evaluated.}
This metric checks whether the final ISPM answer logically follows from the agent’s own 
reasoning trace. An aligned answer must be consistent with SQL evidence, configuration 
enumerations, schema-based interpretations, and identity/privilege paths referenced in 
the reasoning. Misalignment often indicates that the reasoning was ignored or that the 
answer introduces conclusions not justified by evidence.

\paragraph{Criteria.}
\begin{quote}\ttfamily\raggedright
You are a senior expert in the domain `Identity Security Posture Management (ISPM)`, 
specializing in Questions about identities (human and non-human), their privileges, access 
paths, authentication posture (MFA, phishing-resistant methods), lifecycle hygiene 
(onboarding/offboarding, dormant accounts), policy/configuration posture, and resulting 
identity-centric risk across cloud and SaaS environments.

You are evaluating whether an answer is supported by the reasoning trace.

ACTUAL\_OUTPUT is the STEP\_JOURNAL (reasoning trace) for the QUESTION.
CONTEXT includes the final ANSWER and may also include SCHEMAS and RETRIEVED\_EXAMPLES.

Check if the conclusions in the ANSWER logically follow from the reasoning steps and are 
compatible with the schema/domain. Do not require the answer to match historical examples exactly.
\end{quote}

\paragraph{Rubric.}
\begin{quote}\ttfamily\raggedright
0.0 — Misaligned: answer contradicts or ignores major parts of the reasoning.

0.5 — Partially aligned: some answer elements supported, others unclear or unsupported.

1.0 — Fully aligned: answer clearly and logically follows from the reasoning steps.
\end{quote}

\subsection{Contextual Relevancy}

\paragraph{What is evaluated.}
This metric evaluates how relevant the retrieved content and factual evidence are for 
solving the current ISPM question. Two sources matter:  
(1) retrieved historical SQL/QA patterns, and  
(2) factual evidence extracted from the agent’s own step journal.  
High scores indicate that retrieval supports the task and that evidence is appropriate 
for the identity-security reasoning required.

\paragraph{Criteria.}
\begin{quote}\ttfamily\raggedright
You are a senior expert in the domain `Identity Security Posture Management (ISPM)`, 
specializing in Questions about identities (human and non-human), their privileges, 
access paths, authentication posture (MFA, phishing-resistant methods), lifecycle hygiene 
(onboarding/offboarding, dormant accounts), policy/configuration posture, and resulting 
identity-centric risk across cloud and SaaS environments.

You are evaluating the relevancy of context (patterns and factual evidence) retrieved or 
produced during an agent run.

CONTEXT consists of:
  (A) RETRIEVED\_EXAMPLES: historical questions, answers, and/or SQL templates retrieved 
      by a helper agent. These are PATTERNS, not ground-truth facts.
  (B) FACTUAL\_EVIDENCE\_FROM\_STEP\_JOURNAL: telemetry fragments, schema snippets, 
      enumeration outputs, or other factual details extracted from STEP\_JOURNAL that 
      represent real signals discovered during THIS run.

Your task: evaluate how relevant and useful BOTH the patterns (A) and factual evidence (B) 
are for solving the CURRENT QUESTION.

Guidelines:
- For (A), judge PATTERN SIMILARITY and usefulness as templates.
- For (B), judge whether the evidence is actually relevant and informative for answering 
  the question.
- Ignore hallucinated or obviously irrelevant journal artifacts.
- Ignore whether examples match the tenant exactly; focus on their usefulness as patterns.
\end{quote}

\paragraph{Rubric.}
\begin{quote}\ttfamily\raggedright
0.0 — Context (examples/evidence) mostly irrelevant or not useful.

0.5 — Some parts useful, but mixed with irrelevant or weak context.

1.0 — Examples and factual evidence are highly relevant and helpful.
\end{quote}

\subsection{Retrieval Use Quality}

\paragraph{What is evaluated.}
This metric measures whether the agent actually *uses* the retrieved examples and schemas 
effectively while generating SQL or reasoning. It does not judge whether the retrieval 
contents were good — only whether the agent correctly leveraged them. In ISPM questions, 
this corresponds to proper reuse of privilege-analysis templates, IAM lookup patterns, 
and identity-relationship joins.

\paragraph{Criteria.}
\begin{quote}\ttfamily\raggedright
You are a senior expert in the domain `Identity Security Posture Management (ISPM)`, 
specializing in Questions about identities (human and non-human), their privileges, 
access paths, authentication posture (MFA, phishing-resistant methods), lifecycle hygiene 
(onboarding/offboarding, dormant accounts), policy/configuration posture, and resulting 
identity-centric risk across cloud and SaaS environments.

You are evaluating HOW WELL the agent UTILIZES retrieved patterns (examples) and schemas 
when constructing SQL for the current QUESTION.

You will see:
- QUESTION (what needs to be answered).
- ACTUAL\_OUTPUT: the concatenated SQL\_QUERIES generated by the agent.
- CONTEXT, which may include:
  - SCHEMAS: authoritative structures for tables/columns.
  - RETRIEVED\_EXAMPLES: historical questions/answers/SQL patterns (PATTERNS only).
  - STEP\_JOURNAL: reasoning and tool calls that may reference examples.

Your task is to judge UTILIZATION, not mere relevance of the context:
- Did the agent identify and use the MOST RELEVANT examples/patterns?
- Did the SQL structure benefit from these patterns (joins/filters/tables adapted)?
- Did the agent correctly adapt patterns to the CURRENT QUESTION and SCHEMAS?
- Did the agent avoid misusing examples (non-existent columns, wrong joins)?
- Did the agent avoid ignoring obviously useful patterns?

Important distinctions:
- You are NOT judging whether the examples themselves are good.
- You ARE judging how effectively the agent uses what it was given.
\end{quote}

\paragraph{Rubric.}
\begin{quote}\ttfamily\raggedright
0.0 — Poor utilization: SQL ignores helpful patterns OR misuses them (irrelevant templates, 
invalid columns).

0.5 — Partial utilization: some adaptation, but with gaps or misapplications.

1.0 — Good utilization: SQL effectively adapts the most relevant patterns and avoids 
irrelevant or schema-invalid components.
\end{quote}

\subsection{Example Adaptation}

\paragraph{What is evaluated.}
This metric measures whether the agent correctly adapts retrieved SQL or QA templates to 
the current ISPM question. It evaluates: (1) correct pattern selection, and (2) whether 
the modifications reflect the correct schema, privileges, joins, and filters.  
It does NOT check SQL correctness — only adaptation quality.

\paragraph{Criteria.}
\begin{quote}\ttfamily\raggedright
You are a senior expert in the domain `Identity Security Posture Management (ISPM)`, 
specializing in Questions about identities (human and non-human), their privileges, 
access paths, authentication posture (MFA, phishing-resistant methods), lifecycle hygiene 
(onboarding/offboarding, dormant accounts), policy/configuration posture, and resulting 
identity-centric risk across cloud and SaaS environments.

You are evaluating ONLY the agent's ability to ADAPT RETRIEVED EXAMPLES (historical SQL/QA 
patterns) to the CURRENT QUESTION.

You will see:
- QUESTION: the user's task.
- RETRIEVED\_EXAMPLES: historical SQL templates or QA patterns retrieved for this question.
- MODEL\_SQL\_LIST: the SQL generated by the agent.
- SCHEMAS (optional): authoritative structure of tables/columns.

You are evaluating TWO things only:

1) PATTERN IDENTIFICATION:
   - Did the agent pick the MOST RELEVANT examples from the retrieved patterns?
   - Did the agent avoid using irrelevant or misleading examples?

2) PATTERN ADAPTATION:
   - Did the agent correctly MODIFY example templates to fit the CURRENT QUESTION?
   - Did the agent adapt example joins, filters, and table usage to match the SCHEMA?
   - Did the agent drop irrelevant template components when needed?
   - Did the agent integrate multiple example patterns cleanly when appropriate?

Importantly:
- You are NOT judging SQL correctness beyond whether adaptation is valid.
- You are NOT judging retrieval quality.
- You are NOT judging reasoning quality or the final answer.
- You are ONLY judging how well examples were ADAPTED.
\end{quote}

\paragraph{Rubric.}
\begin{quote}\ttfamily\raggedright
0.0 — Poor adaptation: copy-pasted patterns, irrelevant or incorrect examples, schema violations.

0.5 — Partial adaptation: correct ideas but missing modifications or mixing irrelevant elements.

1.0 — Good adaptation: correctly identifies relevant patterns and adapts them cleanly to 
schema and question intent.
\end{quote}

\subsection{SQL Semantic Appropriateness}

\paragraph{What is evaluated.}
This metric assesses whether the generated SQL is structurally and semantically appropriate 
for answering the ISPM question. It evaluates alignment with the schema, logical correctness 
of joins/filters, and whether the plan could meaningfully compute the required identity or 
privilege posture signals. It does NOT require matching ground-truth SQL.

\paragraph{Criteria.}
\begin{quote}\ttfamily\raggedright
You are a senior expert in the domain `Identity Security Posture Management (ISPM)`, 
specializing in Questions about identities (human and non-human), their privileges, 
access paths, authentication posture (MFA, phishing-resistant methods), lifecycle hygiene 
(onboarding/offboarding, dormant accounts), policy/configuration posture, and resulting 
identity-centric risk across cloud and SaaS environments.

You are evaluating the appropriateness of a set of SQL queries for answering a question.

CONTEXT may include SCHEMAS, RETRIEVED\_EXAMPLES (historical SQL patterns, not tenant GT) 
and STEP\_JOURNAL.
ACTUAL\_OUTPUT is the concatenated SQL\_QUERIES for the QUESTION.

Consider whether the SQL aligns with the schemas and represents a sensible pattern-based 
adaptation (if examples are used) to solve the current question.
\end{quote}

\paragraph{Rubric.}
\begin{quote}\ttfamily\raggedright
0.0 — Inappropriate: SQL is illogical, ignores schema, or cannot solve the task.

0.5 — Partially appropriate: some correct concepts, but incomplete or structurally fragile.

1.0 — Appropriate: SQL is well-structured, schema-aligned, and fits the question intent.
\end{quote}

\end{document}